\documentclass[letterpaper, 10 pt, conference]{ieeeconf}  

\IEEEoverridecommandlockouts                              

\usepackage{graphics} 
\usepackage{times} 
\usepackage{amsmath} 
\usepackage{amssymb}  
\usepackage{bbold}
\usepackage{arydshln}
\usepackage{multirow}

\usepackage{xcolor}
\usepackage{soul}

\usepackage[nobiblatex]{xurl}

\usepackage{physics}

\usepackage{pifont}
\newcommand{\cmark}{\ding{51}}
\newcommand{\xmark}{\ding{55}}

\usepackage{adjustbox}

\usepackage{hyperref}

\let\labelindent\IEEElabelindent
\usepackage{enumitem}

\title{\LARGE \bf
Semantic-SuPer: A Semantic-aware Surgical Perception Framework for Endoscopic Tissue Identification, Reconstruction, and Tracking
}

\author{Shan Lin$^1$, Albert J. Miao$^1$, Jingpei Lu$^1$, Shunkai Yu$^1$, Zih-Yun Chiu$^1$, \\Florian Richter$^1$, Michael C. Yip$^1$, \textit{Senior Member, IEEE} 
\thanks{{}$^1$S. Lin, A.J. Miao, J. Lu, S. Yu, Z.Y. Chiu, F. Richter, and M.C. Yip are with the Department of Electrical and Computer Engineering, University of California San Diego, La Jolla, CA 92093, USA.
        (e-mail: \{shl10, amiao, jil360, shyu, zchiu, frichter, yip\}@ucsd.edu) 
This project was funded by the Telemedicine and Advanced Technology Research Center (TATRC) via award MTEC-21-06-MPAI-004 as well as NSF Award \#2045803. 
F. Richter was supported on an NSF Graduate Research Fellowship.%
}}

\begin{document}
\bstctlcite{IEEEexample:BSTcontrol}

\maketitle
\thispagestyle{empty}
\pagestyle{empty}

\begin{abstract} Accurate and robust tracking and reconstruction of the surgical scene is a critical enabling technology toward autonomous robotic surgery. Existing algorithms for 3D perception in surgery mainly rely on geometric information, while we propose to also leverage semantic information inferred from the endoscopic video using image segmentation algorithms. In this paper, we present a novel, comprehensive surgical perception framework, Semantic-SuPer, that integrates geometric and semantic information to facilitate data association, 3D reconstruction, and tracking of endoscopic scenes, benefiting downstream tasks like surgical navigation. The proposed framework is demonstrated on challenging endoscopic data with deforming tissue, showing its advantages over our baseline and several other state-of-the-art approaches. Our code and dataset are available at \href{https://github.com/ucsdarclab/Python-SuPer}{https://github.com/ucsdarclab/Python-SuPer}.

\end{abstract}


\section{Introduction}
With the maturation of surgical robotic platforms like the da Vinci\textregistered{} Surgical System (Intuitive Surgical, Inc., Sunnyvale, CA) well underway, there has been growing interest for these robot platforms to incorporate increased intelligence towards understanding the anatomy of the operative scene. 3D scene understanding of both geometric and semantic features of anatomy enables more effective navigation in patients and is an important step towards the automation of surgical tasks in the future.
Current surgical navigation systems generally utilize segmented preoperative images, such as CT/MRI scans, to provide 3D semantics information. 
Specifically, the 3D semantic segmentation estimated from the preoperative data serves as a map, unchanged throughout the surgery, and matched with the current scene through registration approaches \cite{ieiri2012augmented, metzger2013design, chen2015development, teatini2019effect, zhang2020pathological}. However, whenever there is a large deformation due to surgical operations or different pre- and intra-operative body positioning, the map becomes less reliable, thereby limiting the applications of existing navigation systems to deformable surgical scenes. 

In contrast, video semantic segmentation has the potential to extract precise anatomical information in real-time to update the navigation map.
Existing works have shown the benefits of integrating semantic information extracted from videos with 3D geometry for perception in indoor dynamic environments and autonomous driving \cite{civera2011towards, kundu2014joint, bowman2017probabilistic, zhang2018semantic, chen2019suma++, doherty2020probabilistic, menini2021real, fan2022blitz}. 
Different from environments addressed in these works, surgical scenes usually involve textureless tissue surfaces and constantly deforming tissues. 
These features can cause difficulty in data association, which is a process to find the correspondences between the sensory inputs and the tracked models. Several works have investigated data association approaches, including pixel-level feature matching and dense photometric loss, but still show that this remains a challenge. We aim to alleviate this problem from a different perspective: using semantics to guide data association. We propose a cost function named ``morphing loss" that explicitly encourages the border consistency between the semantic segmentation from the current video frame and the 3D model by projecting its semantic information to the image plane.

\begin{figure}[t]
\centering
\includegraphics[width=0.99\linewidth]{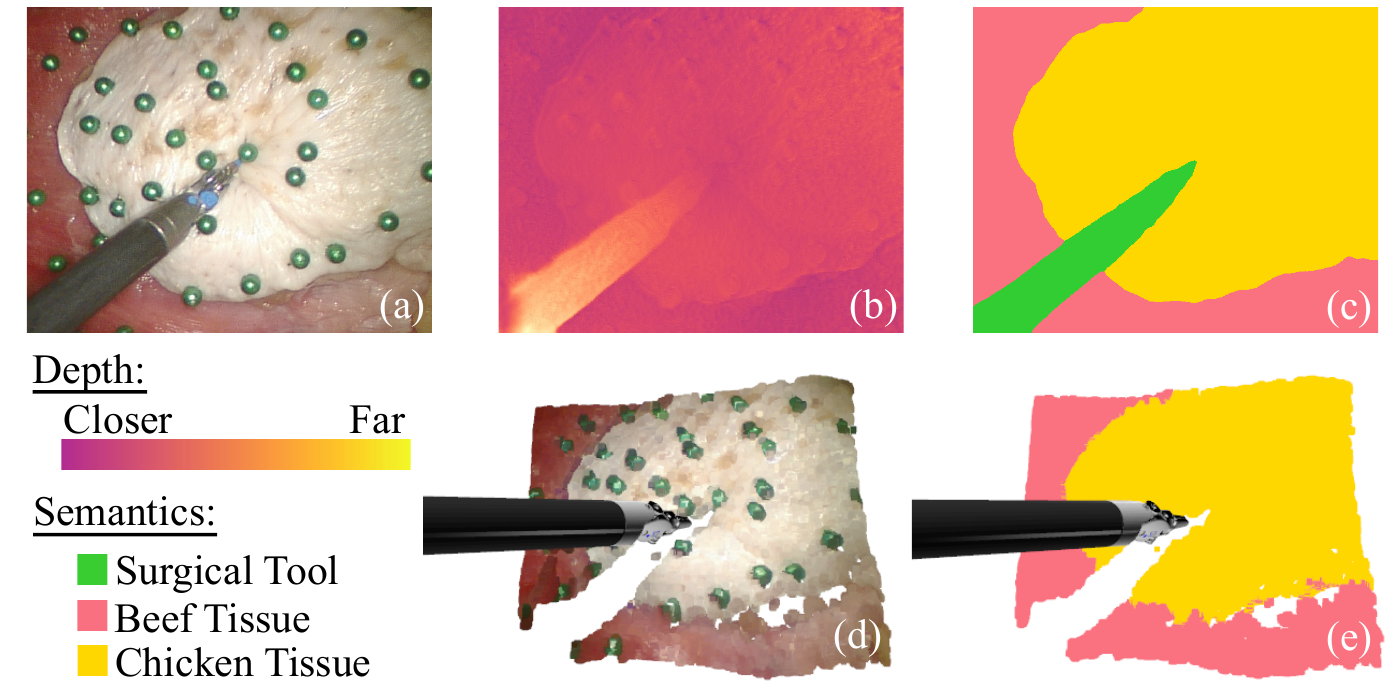}
\vspace{-2em}
\caption{\textbf{A demonstration of Semantic-SuPer.} (a) Input video frame. (b) Depth and (c) semantic segmentation map estimated from the input. (d) Scene rendered from the tracked surfels and surgical tool pose. (e) Scene rendered from surfels visualized by colors corresponding to their semantics.}
\vspace{-1.5em}
\end{figure}

In this work, we propose a novel semantic-aware surgical perception framework, Semantic-SuPer, that combines semantics from 2D videos with 3D deformation tracking and reconstruction. This framework builds upon SuPer \cite{li2020super, lu2021super} by integrating kinematic model-based surgical tool tracking, model-free deformable tissue tracking, and a robotic arm control loop with semantics, increasing robustness and improving accuracy. The main contributions are as follows:
\begin{itemize}
\item A perception framework that integrates semantic information from videos with the 3D geometric properties of the surgical environment for better scene understanding.

\item 
A novel morphing loss that explicitly constrains the semantic consistency between the new input data and the 3D model, showing the benefits of including semantic information for surgical perception.

\item An investigation on the influence of semantic segmentation accuracy on the endoscopic video datasets.

\item A released robotic tissue manipulation dataset for semantic-aware 3D tracking and reconstruction, collected using the da Vinci Research Kit (dVRK) \cite{kazanzides2014open}.
\end{itemize}

\section{Related Work}

Endoscopic tissue deformation tracking and semantic segmentation are two critical components for intelligent assistance in surgery. Because the integration of these two tasks in endoscopic scenes remains under explored, we mainly focus on related work for each individual topic.

\textbf{Surgical Scene Semantic Segmentation} aims to divide a surgical image into regions of different tissues and tools. Motivated by the developments in deep learning, various networks, especially Convolutional Neural Networks (CNNs), have achieved state-of-the-art segmentation performance for endoscopic images \cite{garcia2017toolnet, shvets2018automatic, islam2019learning, allan20192017, qin2020towards, ross2021comparative, lin2021multi}. Most existing works focus on developing segmentation algorithms, while the exploration of their potential applications is still limited. In contrast, we focus on integrating the semantic information with surgical scene tracking in the 3D space.


\textbf{Endoscopic Tissue Tracking}
is a specific area of non-rigid tracking, where the low textured, deformable tissue is a significant challenge. Earlier works typically track the scenes based on rigidness assumptions \cite{grasa2011ekf, grasa2013visual, marmol2019dense}, thus their performance degrades when dealing with large motions. 
MIS-SLAM \cite{song2018mis} described the deformation using embedded deform nodes and achieved better tracking. 
Also, there is another trend of works that model deformations with models like mesh and spline, based on the assumption that the tissue surface is continuous \cite{wong2012quasi, lurie20173d, lamarca2020defslam, gomez2021sd}.
Yet, the complex surgical scenes still make the data association a challenge even for state-of-the-art methods \cite{gomez2021sd, recasens2021endo}, which led to exploration on stronger features as well as photometric losses that impose dense constraints to this process \cite{gomez2021sd, recasens2021endo, yip2012tissue}. Compared to the approaches mentioned above, we aim to build a comprehensive framework that integrates surgical scene tracking and reconstruction with semantic segmentation and show that semantic information can provide guidance for data association.

\textbf{Semantic SLAM} refers to approaches that combine the geometry of the world estimated by SLAM with object detection or semantic segmentation results. Semantic SLAM has been widely studied for areas including autonomous driving, where semantic information has been used to select features or regions of interest, improve data association, identify and remove dynamic points for mapping,  and assist long-term localization \cite{civera2011towards, bowman2017probabilistic, zhang2018semantic, chen2019suma++, doherty2020probabilistic, fan2022blitz}. Yet, methods that combine 3D tracking and reconstruction with semantic segmentation is still very limited for endoscopic data. To the best of our knowledge, there are only two works involving endoscopic data, using a binary mask to segment surgical tools from tissue backgrounds \cite{wu2022semantic, wang2022neural}. In contrast, we consider segmentation information of the whole surgical scene, including different types of tissues.

\section{Methods} \label{ch-label}

\begin{figure*}[!t]
\vspace{0.5em}
\centering
\includegraphics[width=0.99\linewidth]{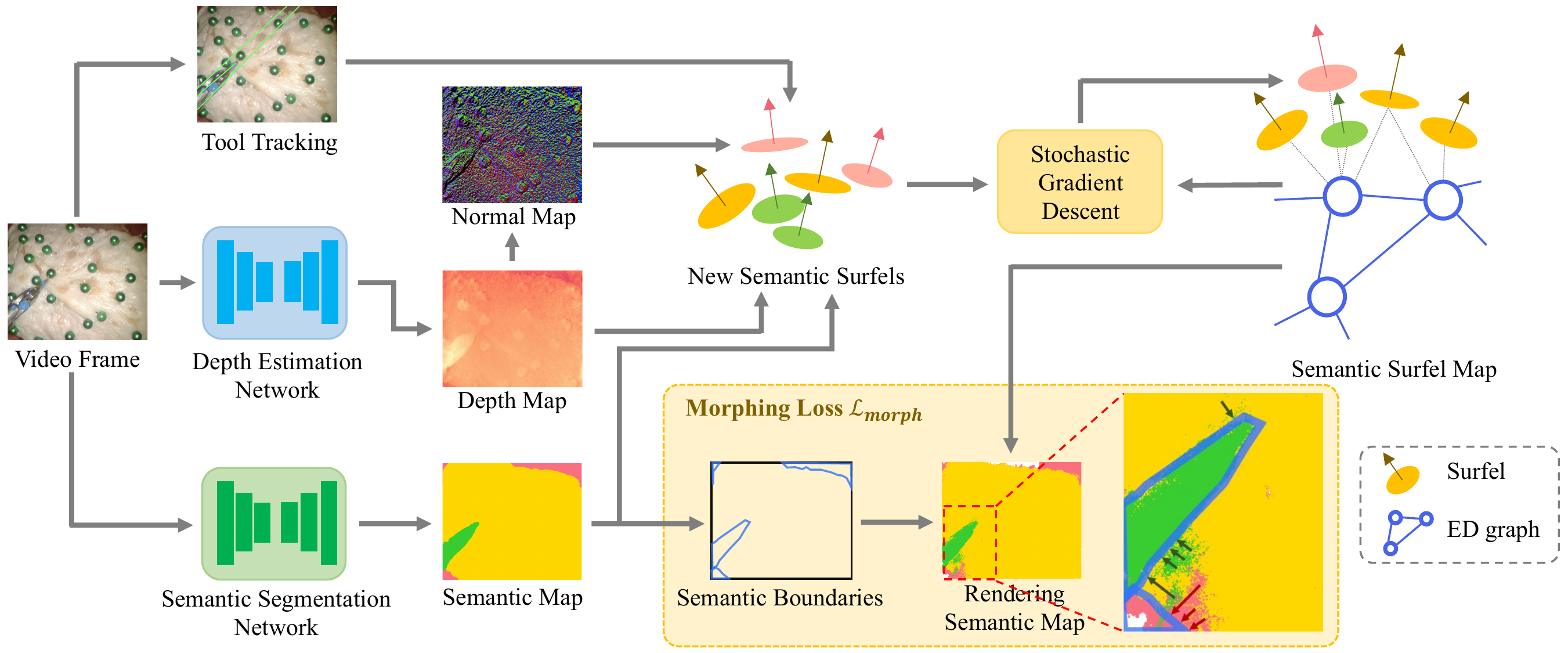}
\vspace{-0.8em}
\caption{\textbf{Overview of the proposed framework.} The depth, semantics, and tool poses are extracted from the input video frame. The transformations of ED nodes are optimized to match the observations under several constraints, including a semantics-based morphing loss. The optimization is implemented using stochastic gradient descent \cite{sratradeoffs} with PyTorch’s automatic differentiation. Surfel position and normal updates are then controlled by the ED nodes.} \label{fig:overview}
\vspace{-1em}
\end{figure*}

The proposed approach builds upon the surgical perception framework SuPer for tissue manipulation \cite{li2020super, lu2021super} by including pixel-wise semantic labels of the endoscopic videos as inputs and building \emph{surface elements (surfels)} \cite{pfister2000surfels, keller2013real, gao2019surfelwarp} with semantic information, as shown in Figure \ref{fig:overview}. We investigate three popular deep segmentation models to provide the semantic labels, while other segmentation architectures can also be used.  
This semantic information is used to suppress erroneous data association between different classes, and therefore improves the robustness of tissue tracking. Furthermore, this allows us to build a 3D surfel map that contains semantic information of different anatomies in deformable surgical scenes, which could benefit endoscopic surgical navigation and other related tasks in the future.

\subsection{SuPer Framework}
SuPer tracks the geometric information of the entire surgical scene, including both the tools controlled by the surgical robot and the deforming tissues. The tool tracking is achieved by a particle filter that utilizes kinematic modeling and information from the endoscopic video. For more details on tool tracking, refer to Richter's \emph{et al.} previous work \cite{9565398}. Meanwhile, the surgical environment (i.e. soft tissue) is tracked using a model-free method based on surfel representation. 
Each surfel $\mathcal{S}$ is defined by a position $\vb{p}_i\in\mathbb{R}^3$, a normal $\vb{n}_i\in\mathbb{R}^3$, a color $\vb{c}_i\in\mathbb{R}^3$, a radius $\mathbb{r}_i\in\mathbb{R}$, a confidence score $\mathbb{c}_i\in\mathbb{R}$, and a time stamp $\mathbb{t}_i\in\mathbb{N}$ of its last update. One can refer to \cite{keller2013real, gao2019surfelwarp} for details on adding, fusing, and deleting surfels.

The number of surfels is proportional to the number of image pixels, so tracking each surfel individually requires a large number of parameters. As inspired by \cite{sumner2007embedded}, SuPer introduces the Embedded Deformation (ED) graph with vertices that are much sparser than the surfels to drive the motion of the surfel set. The ED graph is given by $\mathcal{G}_{ED}=\{\mathcal{V}, \mathcal{E}, \mathcal{P}\}$, where $\mathcal{V}$ is the set of vertices, $\mathcal{E}$ is the set of edges, and $\mathcal{P}$ is the set of parameters. Each vertex (ED node) contains $(\vb{g}_j, \vb{q}_j, \vb{b}_j)\in\mathcal{P}$, where $\vb{g}_j\in\mathbb{R}^3$ is its position, $\vb{q}_j\in\mathbb{R}^4$ and $\vb{b}_j\in\mathbb{R}^3$ are the quaternion and translation parameters, respectively.
The position and normal of each surfel is then updated according to
\begin{equation} \label{update_p}
\widetilde{\overline{\vb{p}}_i} = \vb{T}_g\sum_{j\in \mathcal{N}_i}\omega_j(\vb{p}_i)[T(\vb{q}_j, \vb{b}_j)(\overline{\vb{p}}_i-\vec{\vb{g}}_j)+\vec{\vb{g}}_j]
\end{equation}
\begin{equation} \label{update_n}
\widetilde{\vec{\vb{n}}_i} = \vb{T}_g\sum_{j\in \mathcal{N}_i}\omega_j(\vb{p}_i)[T(\vb{q}_j, 0)\vec{\vb{n}}_i]
\end{equation}
where $\vb{T}_g \in SE(3)$ is the global homogeneous transformation matrix that is shared by all surfels, $T(\cdot) \in SE(3)$ is the local homogeneous transform matrix from a ED node, $\overline{\cdot}$ and $\vec{\cdot}$ are the homogeneous representations of a point and motion (\emph{i.e.} $\overline{\vb{p}}=[\vb{p}, 1]^T$ and $\vec{\vb{g}}=[\vb{g}, 0]^T$), $\mathcal{N}_i$ is the set of $k$-nearest neighbors of $\vb{p}_i$ in $\mathcal{G}_{ED}$. $\omega_j({\vb{p}_i})$ is a weight that indicates the influence of $\vb{g}_i$ to $\vb{p}_i$ and is calculated as $\omega_j(\vb{p}_i)=e^{-\|\vb{p}_i-\vb{g}_j\|}$ and then normalized to sum to one within $\mathcal{N}_i$. (\ref{update_p}) and (\ref{update_n}) can be interpreted as transforming the surfels by averaging the motions of their neighboring ED nodes. 
For every new image frame received, a total of $7\times(n+1)$ parameters ($\vb{q}_j$, $\vb{b}_j$, and $\vb{T}_g$), where $n$ is the total number of ED nodes, will be estimated.
The estimated parameters are then committed (\emph{i.e.}, updating all ED nodes transformations and all surfels based on (\ref{update_p}) and (\ref{update_n})) to track the deformations of the reconstructed tissue.

\subsection{Inputs to the Framework: Depth, Normals, and Semantics} \label{framework_inputs}

As we found previously \cite{li2020super, lu2021super}, the noise in the input depth map is a major factor that limits the performance of the proposed perception framework. Thus, in SuPer-Deep \cite{lu2021super} we did a comprehensive comparison showing that deep learning depth estimation models pre-trained on other datasets, such as the KITTI benchmark \cite{geiger2012we}, led to better tissue tracking performance than traditional stereo matching algorithms. Yet, without finetuning the deep learning models to the surgical dataset, their estimations are still noisy.
For example, the SuPer-Deep framework with PSMNet pretrained on KITTI \cite{chang2018pyramid} crashes within the first 40 image frames from the datasets collected in this work due to the noisy depth observations.
Furthermore, there is not a sufficient amount of surgical data with ground truth depth maps to train supervised stereo matching algorithms.

In this work, we use Monodepth2 which can be trained in a self-supervised fashion \cite{godard2019digging} .
Self-supervision is achieved by maximizing the similarity between the real left image and the reconstructed left image, which is warped from the right view using the estimated depth and camera intrinsics and extrinsics. 
From the depth map, we estimate the surface normal at each pixel as the average of the cross products of all pairs of vectors that point to its 8 neighboring pixels \cite{yang2018unsupervised}.
The surfel set is initialized from the first depth and normal maps.
ED nodes are initialized by sampling uniformly in a rectangle mesh grid in the image, similar to \cite{li2020learning}, and the corresponding node positions $\vb{g}$ are extracted from the point cloud. The ED graph edges are then initialized by connecting each node with its 8 neighbors.


Meanwhile, the deep learning based segmentation model (see Section \ref{sec:seg_methods}) is utilized to predict the segmentation map of each frame. For each surfel, its semantic confidence scores $\vb{s}_i\in\mathbb{R}^C$ ($C$ is the number of classes) are initialized as the corresponding softmax outputs of the network and its semantic label $\mathbb{y}_i\in\mathbb{R}$ is the class that corresponds to the highest softmax score.



\subsection{Semantic-aware Registration} \label{optim}

After the surfels and ED nodes are initialized, they are updated according to new observed depth, normal, and segmentation maps.
The update is applied by solving
\begin{equation}
\label{eq:optimization}
\mathop{\arg \min}\limits_{\vb{q}, \vb{b}, \vb{T}_g} \lambda_s\mathcal{L}_{sim} + \lambda_m\mathcal{L}_{morph} +  \lambda_r\mathcal{L}_{reg}
\vspace{-2mm}
\end{equation}
where 
$\mathcal{L}_{sim}$ the similarity metric that measures the similarity between the transformed model and input data based on data association, $\mathcal{L}_{morph}$ is the proposed semantic-aware morphing loss, $\mathcal{L}_{reg}$ is the regularization term, and $\lambda_{m}, \lambda_{s}, \lambda_{r}$ are hyper-parameters.
The coming subsections detail explicit expressions for each loss term and (\ref{eq:optimization}) is solved in a similar manner as the original SuPer \cite{li2020super}.

\subsubsection{Data Association Approaches}
Data association aims to find the correspondences between the tracked data and new observations (i.e. depth, normal, and semantic maps) which are then used to optimize the transformations.
The first loss is the same point-to-plane ICP loss \cite{low2004linear} as in SuPer \cite{li2020super}:
\begin{equation}
\mathcal{L}_{icp} = \sum_{i} \rho_{i,o} (\vec{\vb{n}}_o^T(\widetilde{\overline{\vb{p}}_i}-\overline{\vb{p}}_o))^2
\vspace{-2mm}
\end{equation}
%
where $\overline{\vb{p}}_o$ and $\vec{\vb{n}}_o$ are the observed positions and normals, bilinearly sampled \cite{jaderberg2015spatial} from the observations at the projected pixel coordinates of $\widetilde{\overline{\vb{p}}_i}$ and $\rho_{i,o}$ is the weight for each term.
The original SuPer uses the naive ICP loss, so $\rho_{i,o} = 1$ \cite{li2020super}.
For Semantic-SuPer, the weight is computed based on the Jensen–Shannon divergence \cite{endres2003new} between the semantic softmax confidences of the surfel and the observation, \emph{i.e.}, $\rho_{i,o} = \exp^{-JSD(\vb{s}_i\|\vb{s}_o)}$. 
The second term utilizes Pulsar \cite{lassner2021pulsar}, a real-time differentiable renderer, to compute a loss directly between a rendering of the tracked soft tissue and the raw image:
\begin{equation}
\mathcal{L}_{render} = \frac{1}{N}\sum_{i=0}^{N}\norm{\frac{1-{SSIM}(I_i, \mathcal{R}(\mathcal{S};\vb{q}, \vb{b}, \vb{T}_g, \vb{K})_i)}{2}}^2
\end{equation}
where ${SSIM}(\cdot)$ is the structural similarity index measure \cite{wang2004image}, $N$ is the number of pixels in the image, $I_i$ denotes the $i$th pixel of image $I$, $\mathcal{R}(\cdot)$ is the Pulsar renderer, and $\vb{K}$ is the camera intrinsic parameters.



\subsubsection{Semantic-aware Morphing Loss}
The data-association losses, $\mathcal{L}_{sim}$, are restricted by the locality of their gradients, which can lead to incorrect associations when the scene is undergoing large deformations.
Therefore, we propose a semantic aware morphing loss that provides longer-range hints by explicitly restricting the semantic edge consistency between the tracked surfel map and every new input semantic map. 
The morphing loss, $\mathcal{L}_{morph}$, minimizes the distance of surfels whose projections onto the image fall outside of their own semantic boundary (see Figure \ref{fig:overview}) and the semantic boundary:
\begin{equation}
    \mathcal{L}_{morph} = \sum_{\pi(\vb{p}_i) \notin \mathcal{R}_i}  \mathop{\min}\limits_{\vb{o} \in \mathcal{B}_i} ||  \pi(\vb{p}_i) - \vb{o} ||^2
    \vspace{-1mm}
\end{equation}
where $\pi(\cdot)$ projects points from the 3D space to the image plane, $\mathcal{R}_i$ is the corresponding semantic region that the 2D projection of $\vb{p}_i$ should lie in, and $\mathcal{B}_i$ is the set of coordinates of boundary pixels of $\mathcal{R}_i$. Note that we minimize the distance in the image plane instead of in the 3D space because the estimated depths around the object boundaries are naturally quite noisy due to the background occlusion~\cite{zhu2020edge}.

\subsubsection{Regularization term}
The regularization term consists of two terms.
The first term, $\mathcal{L}_{face}$, ensures all ED nodes move as rigidly as possible, which helps the model resist incorrect guidance from noisy inputs, and more importantly, provides hints to control some ED nodes that do not receive enough information from observation (\emph{e.g.}, due to occlusion) \cite{sorkine2007rigid, newcombe2015dynamicfusion, li2020super}.
As inspired by \cite{han20202d}, $\mathcal{L}_{face}$ is the $l_2$ norm of the difference between the area of all the transformed triangle surfaces in the ED graph:
\begin{equation}
\mathcal{L}_{face} = \sum_{e_{ij}\in\mathcal{E}, e_{ik}\in\mathcal{E}, j\neq k}\mathbb{I}_{tri}\| \frac{1}{2}|e_{ij}\times e_{ik}| - A_{ijk}\|^2
\end{equation}
where 
$e_{ij}$ denotes the edge that connects the $i$-th and $j$-th ED nodes. 
For each group of three edges that form an triangle, the indicator is set to 1, \emph{i.e.}, $\mathbb{I}_{tri}=1$, and $A_{ijk}$ is the initial area of this triangle; otherwise $\mathbb{I}_{tri}=0$.
The second term, $\mathcal{L}_{Rot}$, is the quaternion normalization term adopted from SuPer to ensure the quaternions hold $\|\vb{q}\|^2=1$:
\begin{equation}
\mathcal{L}_{Rot} = \sum_k \|1-\vb{q}_k^T\vb{q}_k\|^2 \text{ .}
\end{equation}


\section{Experiments and Results}

\subsection{Experimental Setup}

\begin{figure}[t]
\vspace{0.5em}
\centering
\includegraphics[width=\linewidth]{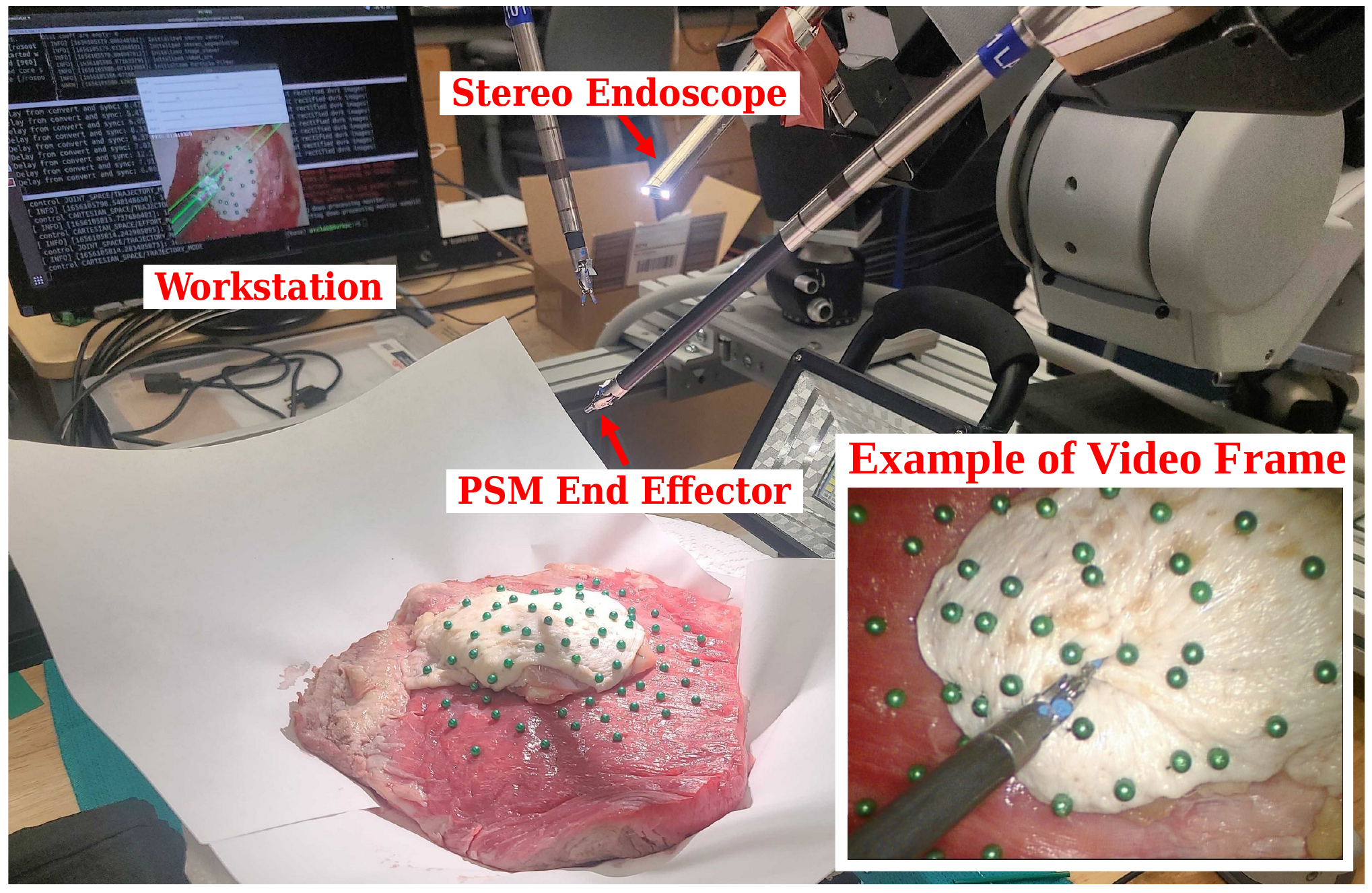}
\vspace{-2em}
\caption{Experimental setup with the dVRK system.}
\label{exp_setup}
\vspace{-1.5em}
\end{figure}
The proposed semantic-aware surgical perception framework was deployed on the da Vinci Research Kit (dVRK) \cite{kazanzides2014open, richter2021bench} for evaluation. 
As shown in Figure \ref{exp_setup}, we overlaid a piece of chicken meat across a piece of beef. The green pins attached on the tissue surface were used to collect ground truth for quantitative evaluation of tracking (Section \ref{metrics}). To generate deformations in the scene, the beef was pushed up-and-down manually from below. Meanwhile, the dVRK was used to control a single surgical robotic arm, \emph{i.e.}, the Patient Side Manipulator (PSM), to grasp and tug the chicken tissue. 
The Robot Operating System (ROS) was used to handle all communication between subsystems, which ran on the Intel i9-7940X and Nvidia RTX 2080 GPU. 4 trials were conducted, each consisting of 150 frames, named Lab 1, 2, 3, and 4 in the following sections. Raw stereo videos were recorded at 1080p, 30 fps, and then rectified to 640$\times$480 resolution. For each trial, timestamped joint angles, gripper poses, and rectified video streams were saved. 

\begin{table*}[t]
\vspace{0.5em}
\caption{Reprojection error comparison on our dataset.}\label{tab:compare_sota}
\vspace{-1.5em}
\begin{center}
\begin{tabular}{lcccc}
\hline
\multirow{2}{*}{Method} & \multicolumn{4}{c}{Data} \\ \cline{2-5}
& Lab1 & Lab2 & Lab3 & Lab4 \\
\hline
DefSLAM \cite{lamarca2020defslam} & 16.5(12.5), 14.5(11.3) & 14.5(13.2), 15.6(13.8) & 12.8(8.8), 13.0(8.5) & 7.0(5.2), 7.3(5.4) \\
SD-DefSLAM \cite{gomez2021sd} & 8.5(8.5), {\bf5.2(4.6)} & {\bf8.0(9.4)}, 10.3(12.0) & 8.4(9.1), 7.1(7.3) & {\bf3.8(4.0)}, {\bf3.1(3.6)}\\
\hline
SuPer \cite{li2020super} & 10.8(8.8), 8.6(7.0) & 10.8(10.1), 10.2(9.1) & 8.1(6.7), 7.2(5.9) & 4.8(4.3), 4.1(3.2) \\
NoSoftLabel-Semantic-SuPer & 12.7(9.5), 9.1(7.4) & 10.8(9.7),10.4(9.6) & 8.9(7.3),7.3(6.0) & 7.1(8.0),13.2(17.6) \\
Semantic-SuPer & {\bf7.5(6.1)}, 6.7(5.7) & 8.6(7.6), {\bf9.2(7.8)} & {\bf6.0(4.9)}, {\bf5.9(4.8)} & 4.3(3.8), 4.3(3.4) \\
\hline
\multicolumn{5}{l}{* From left to right, the two metrics of each data are the reprojection errors averaged over all points and over points near}\\
\multicolumn{5}{l}{object boundaries, respectively. The errors are formatted as `mean(standard deviation)'. The best result in each row is in \textbf{bold}.}\\
\end{tabular}
\end{center}
\vspace{-1em}
\end{table*}

\begin{table*}[t]
\caption{Ablation studies of Semantic-SuPer.} \label{tab:ablation-rst}
\vspace{-1.5em}
\begin{center}
\begin{tabular}{lcccccc}
\hline
\multirow{2}{*}{Method} & \multicolumn{2}{c}{\underline{Cost Functions}} & \multicolumn{4}{c}{Data} \\ \cline{4-7}
& $\mathcal{L}_{morph}$ & $\mathcal{L}_{render}$ & Lab1 & Lab2 & Lab3 & Lab4\\
\hline
\multirow{2}{*}{SuPer\cite{li2020super}} & - & \xmark & 10.8(8.8), 8.6(7.0) & 10.8(10.1), 10.2(9.1) & 8.1(6.7), 7.2(5.9) & 4.8(4.3), 4.1(3.2) \\
& - & \cmark & 10.9(9.1), 8.8(7.1) & 9.9(9.0), {\bf8.7(7.2)} & 8.8(7.2), 8.3(6.9) & 4.7(4.2), 4.0(3.4) \\
\hline
\multirow{4}{*}{Semantic-SuPer} & \xmark & \xmark & 10.0(8.3), 8.5(7.1) & 10.6(9.9), 10.1(9.0) & 8.7(7.1), 8.6(7.1) & 4.7(4.1), 4.0(3.3) \\
& \cmark & \xmark & 7.7(6.3), {\bf6.4(5.4)} & 9.4(8.5), 9.6(8.4) & 6.3(5.0), 6.1(5.0) & 4.5(3.9), 4.4(3.6) \\
& \xmark & \cmark & 10.2(8.3), 9.5(7.5) & 10.2(9.4), 9.4(8.1) & 8.2(6.6), 8.5(7.0) & 4.7(4.1), {\bf3.9(3.1)} \\
& \cmark & \cmark & {\bf7.5(6.1)}, 6.7(5.7) & {\bf8.6(7.6)}, 9.2(7.8) & {\bf6.0(4.9)}, {\bf5.9(4.8)} & {\bf4.3(3.8)}, 4.3(3.4) \\
\hline
\multicolumn{7}{l}{* Refer to the note of Table \ref{tab:compare_sota}.}\\
\end{tabular}
\end{center}
\vspace{-2em}
\end{table*}

\subsection{Metrics and Ground Truth} \label{metrics}

To quantitatively evaluate the tracking results, we attached roughly 60 green pins onto the tissue surface (see Figure \ref{exp_setup}), covering important areas such as those with larger deformations and the boundaries of tissues, and extracted their trajectories in the image plane throughout each trial. The pin trajectories were obtained through a tracking-by-detection paradigm, where detection was achieved by identifying green circular regions in the Hue-Saturation-Value (HSV) color space of the image. Because the videos were captured at different distances to the scene in the 4 trials, the number of green pins that appear in the videos ranges from 30 to 60. The tracked surfels were then projected to the image plane for calculating the reprojection errors, i.e., the distances between the surfel reprojections and their corresponding ground truth.


\subsection{Implementation Details}\label{details}

\subsubsection{Depth Estimation}
Monodepth2 is pre-trained with a larger surgical dataset, the Hamlyn dataset \cite{ye2017self}, and then finetuned with our data. We find that the generalizability of existing self-supervised deep depth estimation models including Monodepth2 is limited when applied to our data, which we believe is because 1) the domain gap between the Hamlyn dataset and our data is not ignorable and we do not have sufficient frames 
for finetuning, and 2) the distances from the scene to the camera vary a lot between different trials. 
Improving depth estimation is not our focus here, so we finetune the model without a train-test split. For both pre-training and finetuning, Monodepth2 is trained for 20 epochs using the Adam optimizer \cite{kingma2014adam}, with a batch size of 16, a learning rate of $10^{-4}$ for the first 15 epochs and $10^{-5}$ for the remainder. The estimated depth map is post-processed by a widely adopted method that fuses the depth maps for the target image and its horizontally flipped image \cite{godard2017unsupervised}.

\subsubsection{Semantic Segmentation} \label{sec:seg_methods}
We compare the influence of three segmentation algorithms, DeepLabv3+ \cite{chen2018encoder}, U-Net \cite{ronneberger2015u}, and UNet++ \cite{zhou2018unet++}, on the performance of Semantic-SuPer. These models are trained for 50 epochs using Adam optimizer, with a batch size of 16 and an initial learning rate of $10^{-4}$ which step decays by 0.1 for every 16 epochs. The models are trained under a K-fold cross-validation setup with $K=4$. At each split, three trials were used to train the model, which was then directly applied to the remaining trial for evaluating Semantic-SuPer.

\subsubsection{Deformable Tracking}
The methods for initializing surfel radius and adding surfels and ED nodes are in the same manner as SuPer \cite{li2020super}.
To initialize the ED graph, since the depths vary a lot between trials, instead of using a fixed step size to generate the mesh, we choose the step size for each trial by ensuring the average edge length of the graph is around 5mm. 
Finally, the hyperparameters that control the cost functions are chosen as $\lambda_m=10$, $\lambda_s=1$, and $\lambda_r=10$.

\begin{figure*}[!t]
\vspace{0.5em}
\centering
\includegraphics[width=0.86\linewidth]{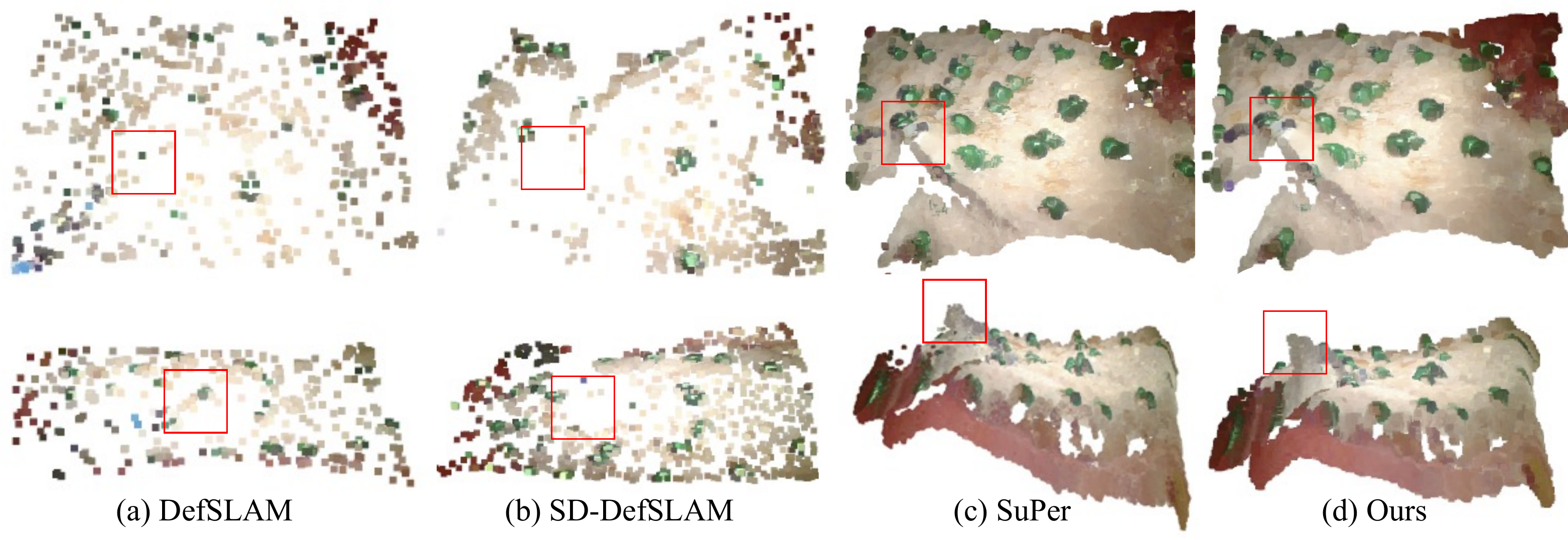}
\vspace{-1em}
\caption{Comparison of the tracked point cloud with SOTA methods. The tissue grasping point is in the red rectangle.}
\label{fig:sample_rst}
\vspace{-1.3em}
\end{figure*}

\begin{figure}[t]
\centering
\includegraphics[width=0.5\textwidth]{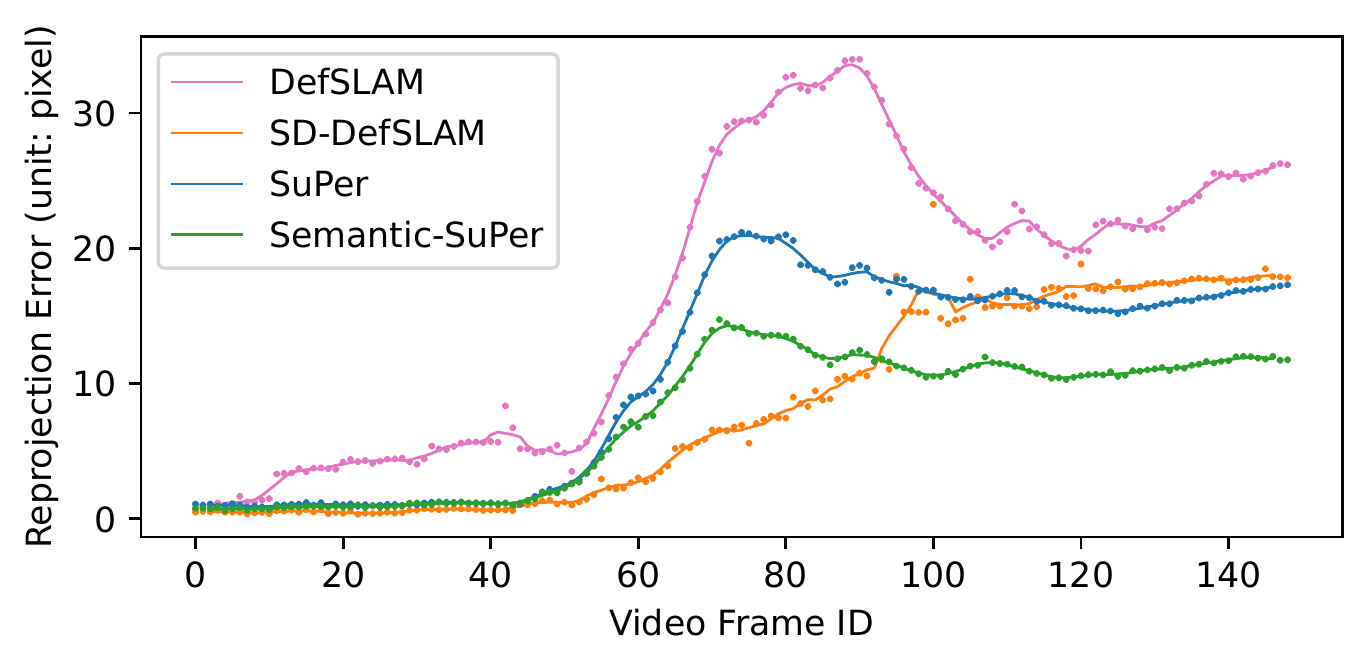}
\vspace{-2.5em}
\caption{Average reprojection error through time in Lab 1.}
\label{fig:quan_plot}
\end{figure}

\subsection{Comparisons with the State-of-the-art Methods}

As shown in Table \ref{tab:compare_sota}, we compare Semantic-SuPer against two baselines: 1) SuPer \cite{li2020super}, and 2) NoSoftLabel-Semantic-SuPer that does not consider the semantic confidence score, only connects surfels and ED nodes that belong to the same class, and uses naive ICP metric calculated from pairs of surfels from the same class. We also evaluate the performance of SOTA deforming surgical scene tracking and reconstruction algorithms DefSLAM \cite{lamarca2020defslam} and SD-DefSLAM \cite{gomez2021sd}. DefSLAM and SD-DefSLAM track the scene based on relatively sparse feature matching and may not track the labeled points, so we estimated the flow of a certain labeled point by averaging the flows of its 3 nearest neighbors. 
We also conduct an ablation study in Table \ref{tab:ablation-rst} to evaluate the effectiveness of the proposed modules on top of SuPer.

Table \ref{tab:compare_sota} shows that Semantic-SuPer outperforms the baselines on Lab 1-3. Lab 4 has the furthest distance between the camera and the scene so it has minimum deformation and changes in the segmentation map, which leads to the small performance gap between Semantic-SuPer and SuPer on it. 
Moreover, our framework outperforms DefSLAM on each of the videos, while achieving either comparable or better performance than SD-DefSLAM. DefSLAM and SD-DefSLAM use matching algorithms based on sparse image features so they could achieve low reprojection errors by selecting more robust features. Yet, because the data was collected by a stationary camera, these two algorithms are unable to reconstruct the 3D surfaces well, while our approach uses monocular depth estimation techniques and thus can provide more accurate and dense tracking, as shown in Figure \ref{fig:sample_rst}.

Figure \ref{fig:quan_plot} shows the average reprojection error of each video frame in Lab 1. The large deformation happens around the 60th to the 110th frames; during this period we can see larger gaps between DefSLAM, SuPer and Semantic-SuPer. SD-DefSLAM has a much parse detection in the region with a larger deformation, \emph{i.e.}, the grasping point shown by the red rectangle in Figure \ref{fig:sample_rst}, so it does not follow the same pattern as the other three methods. It should be noted that the detections near the grasping point are very important because the robot requires this information to perform autonomous manipulation, which is one of our final goals.

\subsection{Influence of different segmentation methods}
A comparison of the influence of different segmentation algorithms on Semantic-SuPer is shown in Table \ref{tab:compare_seg}. We measure the quality of the predicted segmentation maps using Hausdorff distance (HD), which indicates the largest segmentation error, and F\textsubscript{1} score \cite{taha2015metrics}. 
Table \ref{tab:compare_seg} shows a better performance is associated with a better segmentation, \emph{i.e.}, lower HD or higher F1 score. Also, we find that it is more likely to observe bad trackings after a small region within a larger semantic area is incorrectly segmented as another class, which could be addressed by postprocessing the segmentation map using methods such as Conditional Random Fields (CRFs) \cite{zheng2015conditional}.

\begin{table}[t]
\caption{Comparison studies of the influence of segmentation quality on Semantic-SuPer.} \label{tab:compare_seg}
\vspace{-0.5em}
\begin{adjustbox}{width=0.5\textwidth}
\begin{tabular}{lcccc}
\hline
\multirow{2}{*}{Data} & & \multicolumn{3}{c}{Segmentation Method} \\ 
\cline{3-5}
& & DeepLabV3+ & UNet & UNet++ \\
\hline
\multirow{3}{*}{Lab1} & HD(pixel) & {\bf124.6} & 182.7 & 205.3 \\
                      & F\textsubscript{1}(\%) & 96.3  & 96.6  & {\bf96.9} \\
                      & Reproj. Err. & 7.5(6.1), {\bf6.7(5.7)} & 7.3(6.0), 7.2(5.9) & {\bf7.3(5.9)}, 7.4(5.9) \\
\hline
\multirow{3}{*}{Lab2} & HD(pixel)    & {\bf155.6} & 164.9 & 181.9 \\
                      & F\textsubscript{1}(\%)        & 97.2  & 97.3  & {\bf97.6} \\
                      & Reproj. Err. & {\bf8.6(7.6)}, 9.2(7.8) & 11.3(10.9), 12.4(11.5) & 9.0(8.1), {\bf8.5(7.4)} \\
\hline
\multirow{3}{*}{Lab3} & HD(pixel) & {\bf224}   & 369.6 & 313.3 \\
                      & F\textsubscript{1}(\%) & 97.5  & 97.7  & {\bf98.1} \\
                      & Reproj. Err. & {\bf6.0(4.9)}, 5.9(4.8) & 6.2(5.2), 5.7(4.8) & 6.0(5.0), {\bf5.5(4.6)} \\
\hline
\multirow{3}{*}{Lab4} & HD(pixel)    & {\bf107.3} & 156.4 & 175 \\
                      & F\textsubscript{1}(\%)        & 96.1  & {\bf96.8}  & 96.7 \\
                      & Reproj. Err. & {\bf4.3(3.8)}, 4.3(3.4) & {\bf4.3(3.8)}, {\bf3.7(2.9)} & 4.6(3.9), 4.1(3.1) \\
\hline
\multicolumn{5}{l}{* The best result in each row is in \textbf{bold}. }\\
\multicolumn{5}{l}{Refer to Table \ref{tab:compare_sota} for notes on the reprojection errors.}\\
\end{tabular}
\end{adjustbox}
\vspace{-2em}
\end{table}



\section{Discussion}\label{discussion}
Table \ref{tab:compare_sota} demonstrates the benefits of including semantics for tissue tracking, while Table \ref{tab:ablation-rst} shows that the morphing loss augments the effectiveness of ICP and rendering loss. 
Specifically, the rendering loss may not improve ICP-based SuPer / Semantic-SuPer, but when using the morphing loss, the combination of ICP and rendering loss leads to better performance than ICP-based Semantic-SuPer. 
Moreover, it should be noted that the green pins, attached to tissues to allow quantitative evaluation, in fact, make tracking more challenging for our framework. With the pins, the tissue surfaces became uneven, but advanced self-supervised depth estimation methods usually rely on an assumption of the smoothness of the depth and lead to suboptimal depth maps on our data. In addition, we do not use image feature matching methods like the Lucas-Kanade tracker used in SD-DefSLAM \cite{gomez2021sd} for data association and therefore do not take much advantage of the strong features from the green pins. 
Nevertheless, our results are still competitive.

The comparison between Semantic-SuPer and NoSoftLabel-Semantic-SuPer shows the benefits of considering the certainty of segmentation. 
NoSoftLabel-Semantic-SuPer achieves worse performance, because without using the soft semantic labels, the incorrect segmentations assign surfels to ED nodes that belong to other classes, and the estimations of the ED node transformations will be affected more by wrong associations between surfels belonging to different classes. 
Thus, adopting better uncertainty estimation methods for the semantics (also required for depth input) \cite{huang2018efficient, holder2021efficient, poggi2020uncertainty}, as well as leveraging multi-task learning-based cross-task knowledge \cite{zhang2018overview} to estimate uncertainty could lead to better tracking performance.

\section{Conclusions}
In this paper, we present a novel surgical perception framework Semantic-SuPer that achieves better surgical scene 3D reconstruction and tracking by integrating semantic information, showing the benefits of including semantics in this task, which has not been well-explored in prior works. In the future, we will deploy our framework on endoscopic videos captured by moving cameras. We will also investigate multi-task learning to leverage useful information among depth, normal estimation, and semantic segmentation to improve these tasks, whose performance still limits our framework. 
Furthermore, we plan to extract cross-modality knowledge among these framework inputs to achieve better uncertainty estimation for more robust tracking. This perception uncertainty could further augment the combination of advanced control algorithms with our framework, aiming for surgical task automation.




\clearpage
\bibliographystyle{IEEEtran}
\bibliography{IEEEabrv,IEEEexample}

\end{document}